\documentclass[useAMS,usenatbib,usegraphicx]{mn2e}
\usepackage{natbib}
\usepackage{amsmath}
\usepackage{amssymb}
\usepackage{graphicx}
\newcommand{\be}{\begin{equation}}
\newcommand{\ee}{\end{equation}}
\newcommand{\beq}{\begin{eqnarray}}
\newcommand{\eeq}{\end{eqnarray}}
\newcommand{\bear}{\begin{eqnarray}}
\newcommand{\eear}{\end{eqnarray}}
\newcommand{\ba}{\begin{array}}
\newcommand{\ea}{\end{array}}
\title[Constraining the physics of the r-mode instability]{Constraining the physics of the r-mode instability in neutron stars with X-ray and UV observations}
\author[B.Haskell et al.] {Brynmor~Haskell$^{1}$, Nathalie~Degenaar$^{1,2}$, Wynn~C.~G.~Ho$^{3}$\\
$^1$Astronomical Institute ``Anton Pannekoek'', University of Amsterdam, Science Park 904, 1098 XH Amsterdam, Netherlands\\
$^2$Department of Astronomy, University of Michigan, 500 Church Street, Ann Arbor, MI 48109, USA\\
$^3$School of Mathematics, University of Southampton, Southampton SO17 1BJ, UK}

\begin{document}
\maketitle
\begin{abstract} 
Rapidly rotating Neutron Stars in Low Mass X-ray Binaries (LMXBs) may be an interesting source of Gravitational Waves (GWs). In particular, several modes of stellar oscillation may be driven unstable by GW emission, and this can lead to a detectable signal. 
Here we illustrate how current X-ray and ultra-violet (UV) observations can constrain the physics of the r-mode instability. We show that the core temperatures inferred from the data would place many systems well inside the unstable region predicted by standard physical models. However, this is at odds with theoretical expectations. We discuss different mechanisms that could be at work in the stellar interior, and we show how they can modify the instability window and make it consistent with the inferred temperatures.
\end{abstract}


\section{Introduction}

Low Mass X-ray Binaries (LMXBs) were suggested as interesting sources of Gravitational Waves (GWs) more than thirty years ago \citep{PP, Wagoner}. In these systems, a compact object, which in the case of interest is a neutron star (NS), accretes mass from a less evolved low mass companion. The mass donor fills its Roche lobe, and matter is stripped from the outer layers and forms an accretion disc. The disc matter gradually loses angular momentum and spirals in, until it is eventually accreted by the NS. This process leads to angular momentum being transferred to the NS which can then be spun up to millisecond periods in what is know as the ``recycling'' scenario \citep{AlpR,Rad1}

The main reason for invoking GW emission from these systems is the fact that the distribution of spin rates of both LMXBs and millisecond radio pulsars (MSRPs) appears to have a cutoff at around 730 Hz \citep{Chak}, which is well below the centrifugal break up limit \citep{Cook,Haensel}. This observation still holds true today, even as more systems have been added to the sample \citep{Patruno1}.
Thus it is natural to seek a physical mechanism that can prevent NSs from spinning up further. The most obvious candidate is the accretion process itself, as the interaction between the accretion disc and the star can lead to spin equilibrium if the system approaches a propeller phase and further accretion is centrifugally inhibited. This mechanism dictates a correlation between the magnetic field strength and accretion rate \citep{WZ}, a problem which led to the proposal of several GW emission mechanisms that could generate a strong enough torque to set the spin equilibrium of LMXBs \citep{Bildsten98, Nils99, UCB, Cutler}.
Although several authors have reassessed this problem \citep{A1,A2,ABCPHD}, the question remains unresolved and current GW searches are not sensitive enough to give strong constraints \citep{GWs}.

The main GW emission mechanisms that could be at work in accreting systems are ``mountains'', either on the crust \citep{Bildsten98, UCB, mountain} or in the core \citep{OwenQ, cfl}, deformations due to the magnetic field of the star \citep{Cutler,HaskellM, Melatos}, and modes of oscillation of the star being driven unstable and growing to large amplitudes \citep{Nils99}.

We shall focus on the last, specifically the r-mode instability. An r-mode is a toroidal mode of oscillation for which the restoring force is the Coriolis force. It is particularly interesting because it is not only generically unstable to GW emission \citep{Nils98,Morsink} and can thus potentially grow to amplitudes large enough to explain the spin equilibrium of LMXBs, but its modelling requires a detailed understanding of the physics of NS interiors.
The r-mode can grow unstable if GW emission drives it faster than viscosity damps it. This will only happen in a range of temperatures and spin frequencies which depends strongly on the details of the damping mechanisms.
In the standard picture, the main damping agent at low temperatures (below $\approx 10^{10}$ K) is the viscous boundary layer at the crust-core interface \citep{boundary1,boundary2}, while bulk viscosity is the strongest source of damping at high temperatures \citep{review}.
The nature of the damping mechanisms is very sensitive to the interior microphysics and presence of exotica, such as hyperons and deconfined quarks, or large scale superfluid and/or superconducting components \citep{NilsST, Nayyar, hyperon, mannarelli, cfl, lm, rmode, Alf}.  
Furthermore, r-mode oscillations distort the stellar magnetic field, lead to energy dissipation, and possibly prevent the mode from being driven unstable \citep{Luciano}.

In this paper, we examine these mechanisms and compare them to observational constraints on NS spins and temperatures. We use available data on NS surface temperatures from X-ray observations of LMXBs in quiescence and UV observations of millisecond pulsars. We also present new analysis of five systems, which leads to new upper limits on their surface temperatures.
We conclude that the minimal NS model, i.e., that of a star composed of neutrons, protons and electrons (possibly muons) and whose r-mode damping at low temperatures is due to Ekman pumping at the crust-core interface, is not consistent with observations and that additional damping mechanisms are required, unless the r-mode saturates at a very small amplitude. In this case GW emission would not affect the evolution of the system. 

We also discuss additional damping mechanisms that are likely to be at work in NS interiors and may be consistent with observations.

\section{R-mode instability window}
\label{window}
An r-mode is a fluid mode of oscillation of a NS for which the restoring force is the Coriolis force. To leading order in a slow rotation analysis, it is purely toroidal and has the form
\be
\delta \mathbf{v}=\alpha \left(\frac{r}{R}\right)^l R\Omega \mathbf{Y}_{lm}^B \exp{i\omega t},\label{ampio}
\ee
where $\delta \mathbf{v}$ is the Eulerian perturbation of the total fluid velocity,
$\mathbf{Y}_{lm}^B$ is the magnetic-type vector spherical harmonic, $R$ is the stellar radius and $\alpha$ is the (dimensionless) mode amplitude \citep{Owen}. 
The fluid displacement gives rise to a current quadrupole moment and to the emission of GWs, which can drive the mode unstable via the Chandrasekhar-Friedman-Schutz mechanism \citep{Chandrasekhar,Friedman,Nils98,Morsink}.
If GW emission drives the mode growth, then eventually the mode will saturate when energy is transferred to higher order modes due to non-linear couplings. Given the complexity of the full non-linear problem, this process is highly uncertain. Nevertheless, most recent estimates indicate a saturation amplitude $\alpha\approx 10^{-6}-10^{-5}$, \citep{Bondarescu}.  This can be compared to an upper limit of  $\alpha<10^{-4}$ from GW searches conducted with LIGO \citep{Owenr}.
Note that, as we are dealing with a superfluid star, the superfluid neutrons can also flow independently from the charged component (protons and electrons), leading to relative motion. In fact, to second order in the slow rotation analysis, the r-mode will acquire poloidal components along the relative velocity $\delta \mathbf{w}_{\mathrm{p}\mathrm{n}}$ \citep{Dissipo, rmode}.

An r-mode can be driven unstable as long as GW emission drives the oscillation faster than viscosity damps it. This is usually studied in terms of the critical frequency at which the driving and damping timescales are equal.  Solving for the roots of
\be
\frac{1}{\tau_{GW}}=\frac{1}{\tau_V}
\ee
yields an instability curve that depends on frequency and temperature.
$\tau_{GW}$ is the GW driving timescale which (for an $l=m=2$ r-mode and an $n=1$ polytrope) is given by \citep{review}
\be
\tau_{gw}=-47\,{M}_{1.4}^{-1}{R}_{10}^{-4}{P}_{ms}^{6} \mbox{ s},
\ee
with $M_{1.4}$ is the NS mass in units of 1.4 $M_{\odot}$, $R_{10}$ is the NS radius in units of 10 Km and $P_{ms}$ is the NS rotation period in milliseconds.
The viscous damping timescale $\tau_V$ is given by
\be
\frac{1}{\tau_V}=\sum_i\frac{1}{\tau_i},
\ee
where the summation is over the various dissipative channels, labelled with 'i'.
At high temperature (above $\approx 10^{10}$ K) the main contribution is bulk viscosity due to the modified Urca reaction, with a timescale given by \citep{review}
\be
\tau_{BV}=2.7\times 10^{11}\, {M}_{1.4}{R}_{10}^{-1}{P}_{ms}^{2}{T}_9^{-6} \mbox{ s},
\ee
where $T_9$ is the NS {\bf core} temperature in units of $10^9$ K. Note that this form for the bulk viscosity is only appropriate for small perturbations, such that perturbations of the chemical potentials is much smaller than the thermal energy $kT$. For much larger perturbations, the effect of bulk viscosity is significantly stronger, effectively blocking the growth of the r-mode \citep{Alf2}. However, the amplitudes that are necessary for such a scenario are significantly larger than the saturation amplitudes we consider here, so such a possibility will not be discussed further.

\begin{figure}
\centerline{\includegraphics[height=7cm,clip]{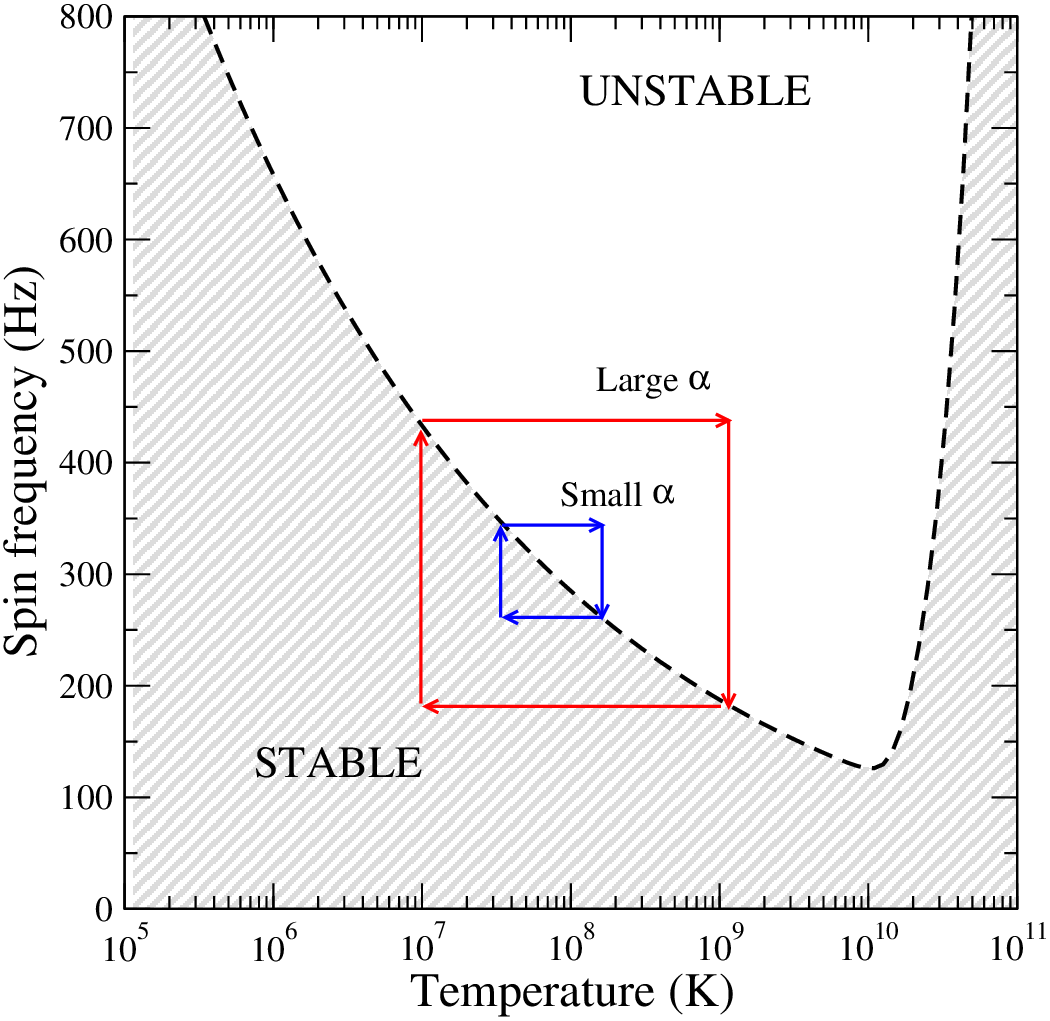}}
\caption{R-mode instability window for the ``minimal'' NS model described in the text, for which the main damping mechanism at low temperature is the Ekman layer at the base of the crust. We schematically illustrate the trajectory a system would follow for high saturation amplitudes ($\alpha\approx 1$) and low saturation amplitudes ($\alpha\approx 10^{-5}$). For large amplitudes, the system undergoes a thermal runaway and heats up significantly but spends much less than $1\%$ of the time in the unstable region. For small saturation amplitudes, the time spent in the unstable region increases, but the spin and temperature variations are modest ($\approx 10\%$; \citealt{Bondarescu}).}
\label{window1}
\end{figure}

At low temperatures, the main source of damping is the viscous boundary layer at the crust-core interface, which leads to a damping timescale
\be
\tau_{EK}=3\times 10^5\,{P}_{ms}^{1/2}{T}_{9} \mbox{ s},
\ee
where we use the estimate of \citet{Ek1} with a ``slippage'' factor $\mathcal{S}=0.05$. The slippage factor accounts for the fact that the crust will not be completely rigid, but will also participate in the oscillation. It is essentially the ratio between the crust/core velocity difference and the mode velocity, so that $\mathcal{S}=1$ corresponds to a completely rigid crust, while smaller values indicate that the mode can penetrate the crust to some extent. Shear viscosity will also play a role at low temperatures, but its effect will be weaker than that of the crust-core interface \citep{review}; thus we neglect it here.
In Figure~\ref{window1}, we show an example of an instability window for the ``minimal'' NS model described above and schematically illustrate the trajectory that an LMXB would follow. For large saturation amplitudes ($\alpha >10^{-3}$), the system undergoes a thermal runaway, i.e., heating up rapidly due to the unstable mode and then spinning down due to the emission of GWs \citep{Levincycle}. It would then enter the stable region, cool and spin up again, closing the cycle. For such a scenario, the system would spend less than $1\%$ of the time in the unstable region, making it very unlikely to observe a system in this stage \citep{Heyl02}. For more realistic values of the saturation amplitude ($\alpha\approx 10^{-5}$), the system could spend up to $30\%$ of the time in the unstable region. The excursion in temperature and frequency are expected to be modest and one would still not expect to see a system well inside the instability window \citep{Bondarescu}.
A further possibility is that the saturation amplitude is very small (which could be the case for low viscosity) and thus the GW torque is very weak. In this case, the GW torque cannot counteract the accretion torque, and a system would reach an equilibrium between r-mode heating and neutrino cooling, while spinning up into the instability window \citep{Bondarescu}.

The minimal cooling model described above is modified if there are hyperons in the NS core. The bulk viscosity can then be much stronger at lower temperatures.
% The problem is complicated by the fact that the damping timescale has a minimum in the temperature range of interest and the timescale does not lend itself to a simple parametrisation.
The effect of hyperon bulk viscosity  on the r-mode instability window has been studied in detail for superfluid NSs by \citet{hyperon}, whose results we shall use below.

Furthermore if, as is generally believed, the core of the NS contains large scale superfluid components, these will rotate by forming an array of quantised vortices. The interaction of vortices with the charged components gives rise to a dissipative force known as mutual friction.  For temperatures well-below the superfluid transition temperature, the damping timescale for mutual friction is roughly constant, but the timescale can vary considerably when the temperature is near the transition temperature. Here we shall use the detailed results of \citet{rmode}.
The main microphysical input that is needed to calculate the mutual friction damping timescale is the value of the (dimensionless) drag parameter $\mathcal{R}$. It has been shown by \citet{lm} and \citet{rmode} that the standard drag parameter (describing electron scattering off vortex cores; $\mathcal{R}\approx 10^{-4}$) does not significantly affect the instability window. However, the situation may be considerably different if the core of the NS is in a type II superconducting state. In this case, the magnetic field is arranged in flux tubes, and their interaction with neutron vortices could lead to strong dissipation, with drag parameters possibly of the order of $\mathcal{R}\approx 10^{-2}$ \citep{flux2, super, flux3}.

Finally it should be noted that, in magnetised stars, fluid motion distorts magnetic field lines, possibly leading to energy being drawn from the mode faster than GW emission can drive it \citep{Luciano}.

\section{Neutron star temperatures and spin rates}

As is obvious from the discussion in Section~\ref{window}, if we wish to construct an instability curve in the frequency versus temperature plane, it is necessary to estimate the temperature of the NS core, on which the damping timescales will depend.
This is clearly not a straightforward task, as what is measured is the surface emission as detected by a distant observer.
In order to estimate the core temperature, we shall use X-ray observations of LMXBs in quiescence (when most of the thermal emission is thought to come directly from the NS surface; see, e.g., \citealt{B98}) and the few available millisecond radio pulsar thermal spectra observed in UV.
This is in contrast to the estimates made by \citet{windowW}, which made use of X-ray observations of LMXBs during bursts.

Several LMXBs have surface temperatures obtained from blackbody fits to their observed X-ray spectrum. For others, the spectrum is completely non-thermal, and only upper limits on the temperature can be obtained. In table 1, we list LMXBs that have a measured temperature (or upper limit) and spin rate. The spin rates are either measured directly for those NSs that display coherent X-ray pulsations (indicated as ``accretion powered'') or inferred from the frequency of oscillations seen during thermonuclear type-I X-ray bursts (labelled as ``nuclear powered''). The spin rates are taken from the overview given by \citet{Patruno1}. For the temperatures, we use the overview compiled by \citet{Heinke1,Heinke2} and include 10 additional sources reported in the literature or analyzed in this work (see Section~3.1). We also include three millisecond radio pulsars for which the temperature was constrained by fitting the UV spectrum.

\begin{table*}
\begin{minipage}{150mm}
\caption{Surface temperatures and spin rates for LMXBs that have measurements (or upper limits) of both. Note that $T_\infty$ is related to the surface temperature $T_S$ by the relation $T_s=(1+z)T_\infty$ where $1+z=(1-2GM/Rc^2)^{-1/2}$ is the red-shift factor. We take $1+z=1.3$ and the core temperature is then obtained as described in the text. We include three millisecond pulsars that have estimates of the surface temperature from UV observations. We distinguish between accretion powered (AP) pulsars (which are observed as X-ray pulsars), nuclear powered (NP) pulsars (for which the spin rate is estimated from type I X-ray burst oscillations) and radio pulsars (RPs). Pulsars that show both burst oscillations and X-ray pulsations are classified as accretion powered. Spin frequencies for the accreting systems are taken from \citet{Patruno1}.
% We also list the core temperature that we infer in this paper.}
}
\begin{flushleft}
\begin{tabular}{l l l l l l l l l l }
Source& & $\nu$ (Hz)& & T$_\infty/10^6$ K& T$_{\mbox{core}}/10^8$ K& &Type& &Reference\\
\hline
Aql X-1& &550& &1.09& 1.08& & AP& &\citet{Heinke1}\\
4U 1608-52& &620& & 1.97& 4.55& & NP& &\citet{Heinke1}\\
KS 1731-260& & 526& & 0.73&0.42& & NP& &\citet{Cackett10}\\
MXB 1659-298& &556& &0.63&0.31& & NP& &\citet{Cackett08}\\
SAX J1748.9-2021& &442& &1.01&0.89& & AP& &\citet{Heinke1}\\
IGR 00291+5934& &599& &0.82&0.54& & AP& &\citet{Heinke2}\\
SAX J1808.4-3658& &401& & $<$0.35&$<$0.11& & AP& &\citet{Heinke2}\\
XTE J1751-305& &435& &$<$0.82&$<$0.54& & AP& &\citet{Heinke2}\\
XTE J0929-314& &185& &$<$0.58&$<$0.26& & AP& &\citet{Heinke2}\\
XTE J1807-294& &190& &$<$0.59&$<$0.27& & AP& &\citet{Heinke2}\\
XTE J1814-338& &314& &$<$0.80&$<$0.51& & AP& &\citet{Heinke2}\\
EXO 0748-676& &552& &1.26&1.58& & NP& &\citet{Deg11T}\\
HETE J1900.1-2455& &377& &$<$0.65&$<$0.33& & AP& &This work\\
IGR J17191-2821& &294& &$<$0.86&$<$0.60& & NP& &This work\\
{IGR J17511-3057}& &245& &$<$1.10&$<$1.10& & AP& &This work\\
SAX J1750.8-2900& & 601& &1.72&3.38& & NP& &\citet{Lowell}\\
NGC 6440 X-2& &205& &$<$0.37&$<$ 0.12& & AP& &This work\\
SWIFT J1756-2508& &182& &$<$0.96&$<$ 0.78& & AP& &This work\\
{SWIFT J1749.4-2807}& &518& &$<${ 1.27}&$<$ 1.61& & AP& &\citet{Degprep}\\
\hline
J0437-4715& &174& &0.12&0.018& & RP& & \citet{UV}\\
J2124-3358& &203& &$<$0.46&$<$0.17&  & RP& &\citet{UV}\\
J0030+0451& &205& &$<$0.92&$<$0.70&  & RP & &\citet{UV}\\
\end{tabular}\label{tempspin}
\end{flushleft}
\end{minipage}
\end{table*}

\subsection{Observations}
In order to obtain constraints on the surface temperature for some sources,
% For several NS with known spin periods, it is possible to put upper limits on their quiescent thermal emission from X-ray observations. With those constraints at hand, we can deduce upper limits on the NS surface temperature, as seen by an observer at infinity, by
we first simulate a fiducial X-ray spectrum with the software package XSpec [v 12.6; \citep{xspec}]. This is done using the NS atmosphere model NSATMOS \citep{Heinke2006}, where we take $M=1.4~M_\odot$ and $R=10$~km. We assume the entire NS surface is emitting (i.e., model normalization is fixed to 1) and use source distances reported in the literature (see below). After constructing such a model for each source, we determine the NS temperature that produces the observed (quiescent) thermal flux limit. This value is then considered to be the upper limit on the NS surface temperature $T_s$.

We use flux upper limits reported in the literature to infer constraints on the surface temperature for three sources: IGR J17191-2821 ($D=11$ kpc; \citealt{Alta10}), NGC X-2 ($D=8.5$ kpc; \citealt{Heinke2010}) and Swift J1756-2508 ($D=8$ kpc; \citealt{Patruno2010}). In the case of IGR J17511-3057, nothing is reported in the literature about its quiescent properties. However, we found two observations obtained with the X-ray Telescope (XRT) onboard Swift, which did not reveal the source during its quiescent state. We obtain an upper limit on the 0.5--10 keV unabsorbed flux of $\sim 7.5 \times 10^{-13}~\mathrm{erg~cm}^{-2}~\mathrm{s}^{-1}$.  We infer a NS surface temperature of $T_s < 1.44 \times 10^{6}$~K for the fiducial model parameters mentioned above and assuming $D=6.9$ kpc \citep{Alta2}.
We analysed a recent XMM-Newton observation of Swift J1749.4-2807 during quiescence (Degenaar et al. in preparation). The source is clearly detected during the observation, but its X-ray spectrum is completely non-thermal. For a distance of $D=6.7$~kpc \citep{Altamirano2011}, we obtain an upper limit on the NS surface temperature of $T_s < 1.66 \times 10^{6}$~K. Finally, HETE J1900.1-2455 has been continuously active since its discovery in 2005, but the source intensity dropped dramatically during a short $\sim 20$-day interval in 2007 \citep{decline,non_det}.  At a certain point the source could not be detected with Swift/XRT; this resulted in an upper limit on the quiescent X-ray flux \citep{non_det}. We re-analysed the data of this non-detection to estimate the upper limit on the NS surface temperature, where we assumed a source distance of $D=3.6$~kpc \citep{Galloway}.

\subsection{Neutron star core temperatures}

Having determined NS surface temperatures, we now estimate the core temperatures.
We assume that the core and crust are nearly isothermal (which is very nearly the case since the thermal conductivity of the crust is high, as indicated by recent cooling observations of X-ray transients; \citealt{BrownN}) and that the core temperature is simply the temperature at the base of the heat blanketing envelope. As we are considering accreting systems and radio pulsars that are thought to have been recycled through accretion, we use the relation between surface temperature and envelope base temperature for a partially accreted crust given by \citet{Pot}, where we follow \citet{BrownN} by considering a layer of light elements down to a column depth of $P/g=10^9$ g/cm$^{-2}$.

We assume that millisecond radio pulsars have been recycled to rapid rotation by accretion and that they have similar crustal compositions as the LMXBs; thus we again use the relation of \citet{Pot} to estimate their core temperature. This is of course a crude assumption. However by using the iron envelope relation of \citet{Gudmundsson}, the estimated temperatures change by a factor of approximately 2. As we shall see, the MSRPs fall in a region of the instability window for which such a correction has no effect on our conclusions; this justifies our use of the relation from \citet{Pot}.
Let us remark that the estimates of core temperature have large uncertainties, as not only is the composition of the envelope uncertain, but some systems may still be thermally relaxing to a steady state after an outburst and may have sizable temperature gradients in the crust. These effects lead to an uncertainty of a factor of a few in the inferred core temperatures \citep{BrownN}. This has no qualitative impact on our conclusions, but this uncertainty should be kept in mind and we shall attempt to quantify it in the following sections.

\subsection{Strange stars}
\label{quark}

It is possible that the most stable form of matter at the high densities that characterise a NS interior may be that of a conglomerate of deconfined quarks \citep{Itoh, Witten}. In fact, it has been suggested that all NSs may be strange stars \citep{Olinto}. Although the ground state of matter at asymptotically high densities and low temperatures is known to be given by paired quarks in the so-called ``Colour Flavour Locked'' (CFL) phase \citep{Alford}, the properties of matter at realistic NS densities are still uncertain. The effect of a CFL core on the instability window was calculated by \citet{cfl} and found to be quite weak, so we shall not consider this possibility any further. We shall thus consider the effect of a conglomerate of unpaired quarks in the NS interior on the r-mode instability window, as this can be quite significant. 
In this scenario the shear viscosity dissipation timescale for a strange star (again assuming an $n=1$ polytrope as the background model) is found to be \citep{NilsST}
\be
t_s\approx 7.4\times 10^7 \left(\frac{\alpha_S}{0.1}\right)^{5/3}{M}_{1.4}^{-5/9}{R}_{10}^{11/3}{T}_{9}^{5/3} \mbox{ s},
\ee
where $\alpha_S$ is the strong coupling constant. 
For the temperature range of interest (below $\approx 10^9$ K), the bulk viscosity damping timescale is \citep{Madsen, NilsST}
\be
t_{Bv}\approx 7.9\left(\frac{m_s}{100 \mbox{MeV}}\right)^{-4}{M}_{1.4}^{2}{R}_{10}^{-4}{T}_{9}^{-2}{P}_{ms}^{2} \mbox{ s},
\ee
where $m_s$ is the mass of the strange quark. A strange star can also support a thin crust of normal nuclear matter up to the neutron drip density, after which free neutrons will drip into the strange core. Such a crust would obviously be much thinner and much less massive than that of a NS, with a maximum mass of approximately $10^{-5}$ M$_{\odot}$ \citep{GlenWeb}. This crust would be much less rigid than a standard NS crust and not contribute significantly to the damping \citep{NilsST}. We shall thus not consider damping due to the crust-core interface in the discussion on strange stars, although its inclusion would not qualitatively change our conclusions.

The presence of the crust is, however, very significant for estimating the temperature of the core, as it provides a ``heat blanket'' for the strange core, allowing the outgoing radiation to thermalise (which would not be the case for a bare strange star, for which the spectrum would be considerably harder; \citealt{PageST}). Furthermore, many of the systems we consider show not only coherent pulsations but also thermonuclear bursts, which would be challenging to explain if there is no crust of normal matter. We shall thus use the same prescription as in the NS case to estimate the core temperature (see also the discussion in \citealt{PierreST}).

\section{Observational constraints on the instability window}

Let us now examine how the observational evidence compares with theoretical calculations of the r-mode instability window. First of all, we begin by comparing the measured temperatures/spins with the minimal model instability window of Section~\ref{window}. 
\begin{figure*}
\begin{minipage}{150mm}
\centerline{\includegraphics[height=7cm,clip]{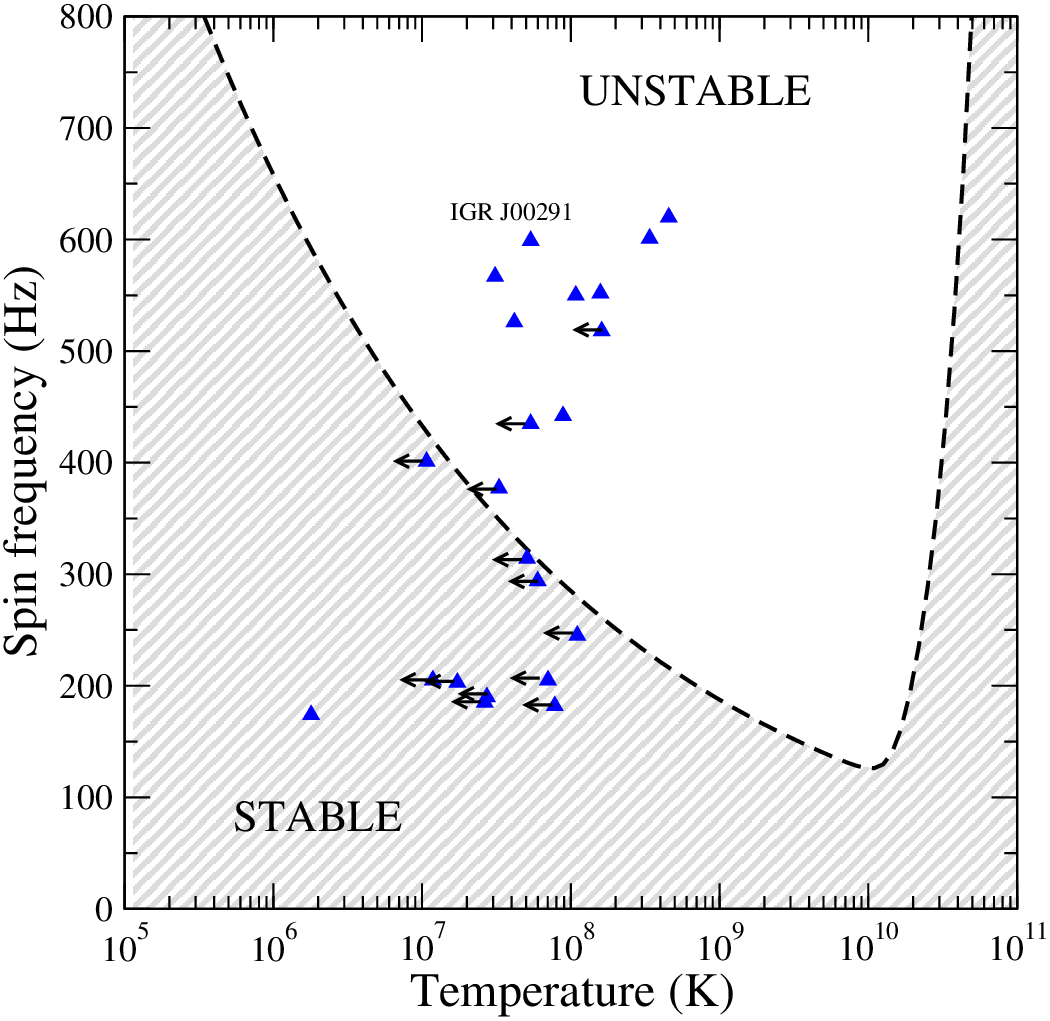}\includegraphics[height=7cm,clip]{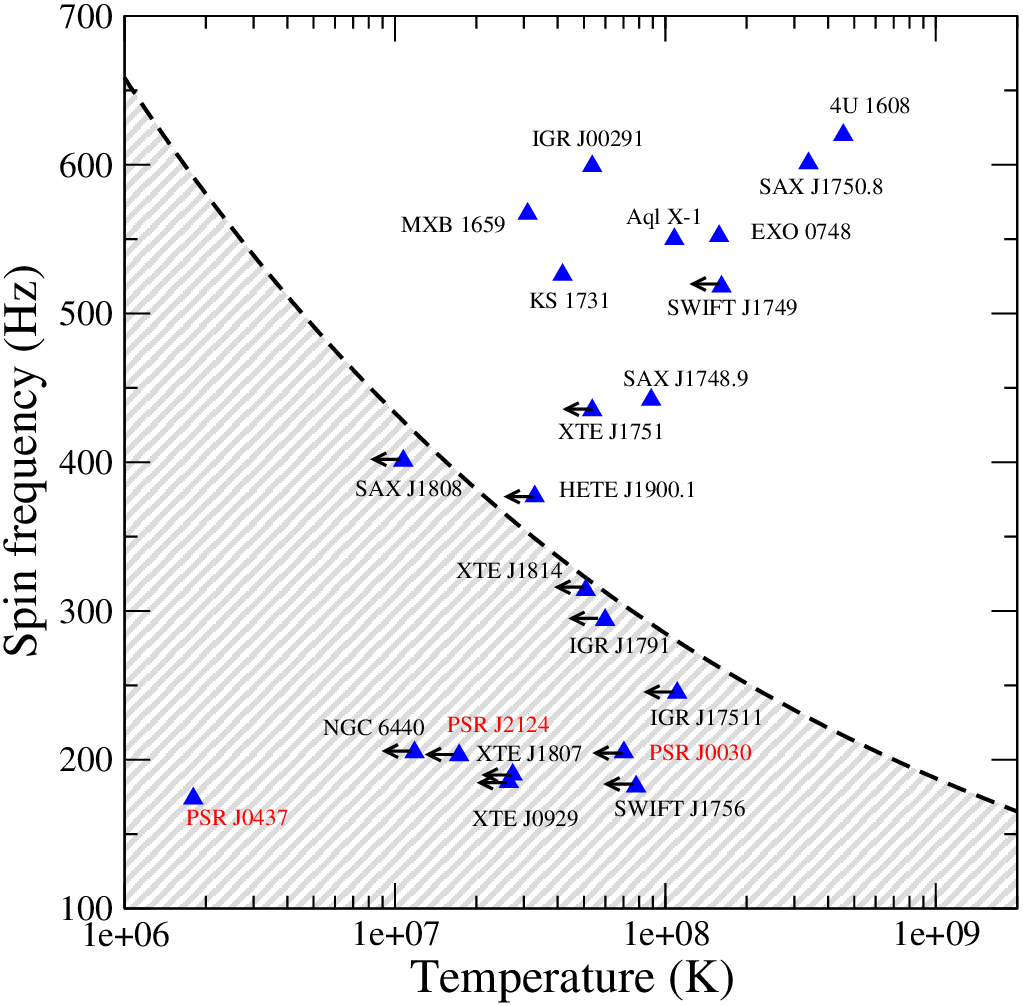}}
\caption{R-mode instability window of LMXBs and MSRPs that have estimates of both the spin frequency and surface temperature (arrows indicate upper limits).  The right panel is the same as the left panel but focused on the low temperature region in which the observed systems are located.  It is obvious that a significant number of systems is well inside the ``minimal'' instability window, where one would not expect to find so many systems. In fact, for realistic values of the saturation amplitude, a star could not heat up enough to be significantly inside the unstable region, while for high values of the saturation amplitude a system would spend only a very small fraction of the time (less than $1\%$) above the instability curve, making it very unlikely to catch systems in this region.  The only possibilities are thus that either the instability curve is significantly different from our minimal model curve due to additional damping mechanisms or the saturation amplitude is small enough not to affect the evolution of the systems.}
\label{observe1}
\end{minipage}
\end{figure*} 
In Figure \ref{observe1} we show the inferred core temperatures and in Figure \ref{errors} we estimate the uncertainty due to the modelling of the outer layers of the star. The error bars on the core temperatures inferred from observations have been obtained by considering two extreme compositions for the stellar envelope, the properties of which (composition, thermal conductivity etc...) control the heat flow from the core to the exterior. We have thus calculated a ``minimum'' temperature by assuming a completely accreted crust of light elements \citep{Pot} and a ``maximum'' temperature by assuming an iron envelope \citep{Gudmundsson}.
As we can see this produces an uncertainty of a factor of a few, which will dominate over the observational uncertainty but does not affect our conclusions.
It is obvious from Figures~\ref{observe1} and \ref{errors} that, even accounting for theoretical and observational uncertainties associated with temperature measurements, several systems are well inside the unstable region. As already mentioned, this would is possible if the saturation amplitude is very large and one is lucky enough to catch the system while it is still in the GW emitting part of its duty cycle. However, given that for large amplitudes ($\alpha > 10^{-3} $) one would expect the system to spend less than $1\%$ of the time in the unstable region, it is highly unlikely that we are observing so many systems in this phase. Furthermore, such a system would be spinning down rapidly due to the emission of GWs, but one of the systems, IGR J00291+5934, has a measured spin-down rate in quiescence of $\dot{\nu}\approx 3\times 10^{-15}$ Hz s$^{-1}$ \citep{Patruno1, Jake, IGR2}, which is consistent with purely electromagnetic spin-down due to a $B\approx 10^8$ G magnetic field (although one cannot rule out a much weaker magnetic field and low level GW emission, see \citet{BrynAle} for a discussion of why this is unlikely to be the case in two other sources, SAX J1808.4-3658 and XTE 1814-338).
A final possibility is that systems inside the window have undergone a thermal runaway and have reached an equilibrium between heating and cooling \citep{Bondarescu}; they are now either at spin equilibrium (i.e., with the GW spin-down torque balancing the accretion spin-up torque) or approaching spin equilibrium (as could be the case for IGR J00291+5934 which exhibits long-term spin-up). We discuss this possibility further in the following section.
It is, however, clear that the minimal model is not consistent with observations.

\begin{figure}
\centerline{\includegraphics[height=7cm,clip]{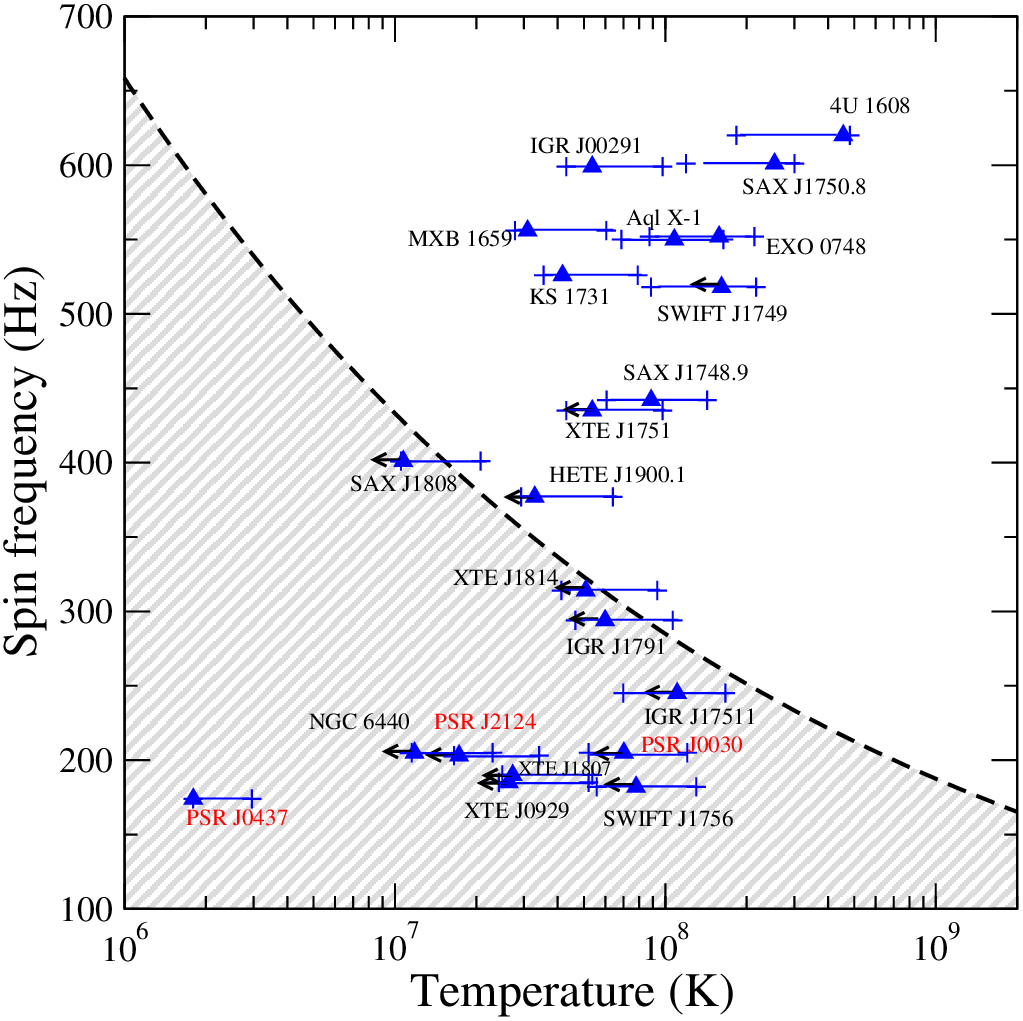}}
\caption{The same r-mode instability window as in the right panel of Figure \ref{observe1} where we have also estimated the error bars due to the uncertainty in modelling the outer layers of the NS, as described in the text. We can see that although there is a significant uncertainty on the inferred core temperatures it is not large enough to modify the conclusion that many of the systems appear to be well inside the unstable region.}
\label{errors}
\end{figure}
\begin{figure}
\centerline{\includegraphics[height=7cm,clip]{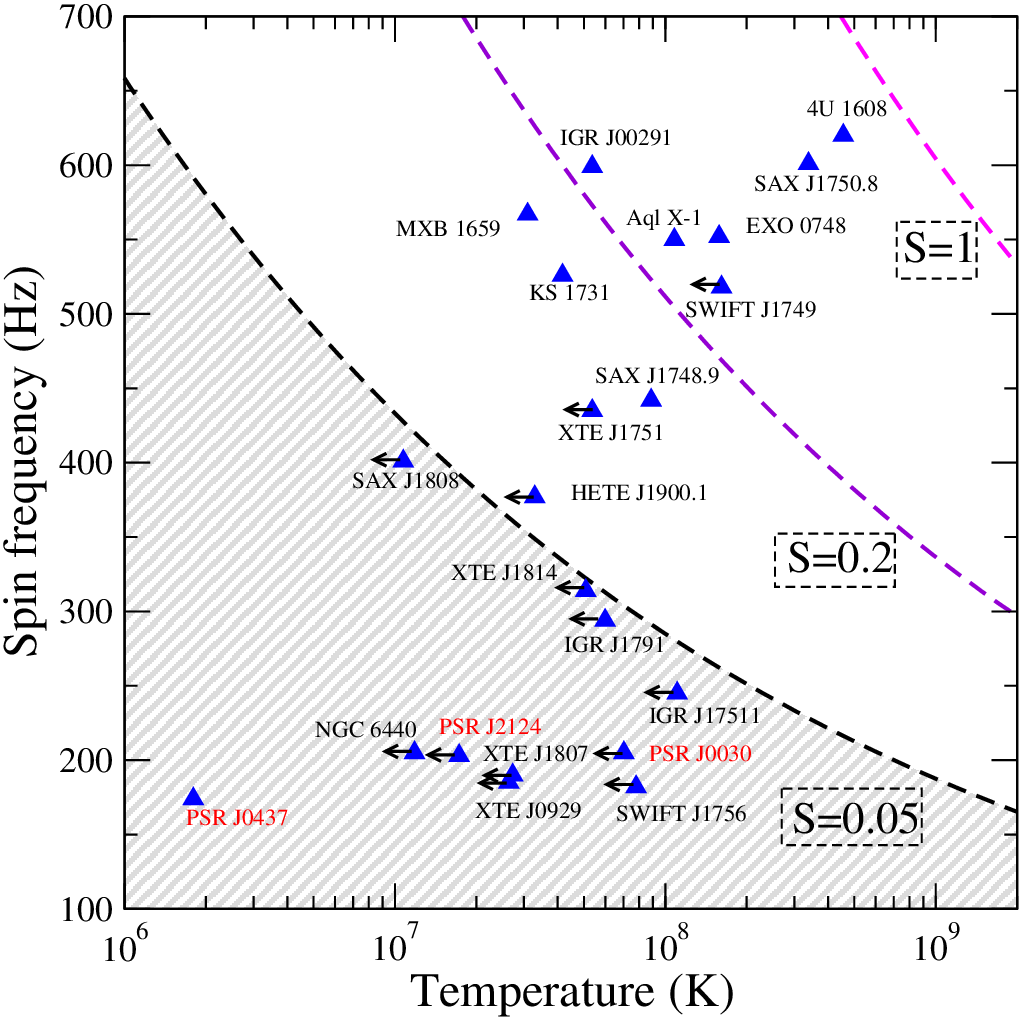}}
\caption{R-mode instability window for different values of the ``slip'' parameter $\mathcal{S}$ (\citealt{Ek1}; see text). A large slip parameter, corresponding to a nearly completely rigid crust, appears to be necessary to explain the observations.}
\label{slip}
\end{figure}
We now discuss the possible mechanisms that may be at work in a realistic NS and that could be consistent with observations.
We first examine effects due to properties of the crust.  One is that the crust may be more rigid than is commonly assumed. This would lead to stronger dissipation at the crust-core interface. In Figure~\ref{slip}, we show the effect of increasing the ``slippage'' factor $\mathcal{S}$ from a standard value of $\mathcal{S}=0.05$ \citep{Ek1} to $\mathcal{S}=1$ (a completely rigid crust). It is obvious that a more rigid crust could allow all the systems to be stable (see also \citealt{Wen}). However such a rigid ($\mathcal{S}=1$) crust is not realistic. Alternatively, given the frequency range of r-modes, the mode may couple effectively to torsional oscillations of the crust. This would produce strong dissipation at the resonance frequency \citep{boundary2,Ek1, windowW}

\begin{figure}
\centerline{\includegraphics[height=7cm,clip]{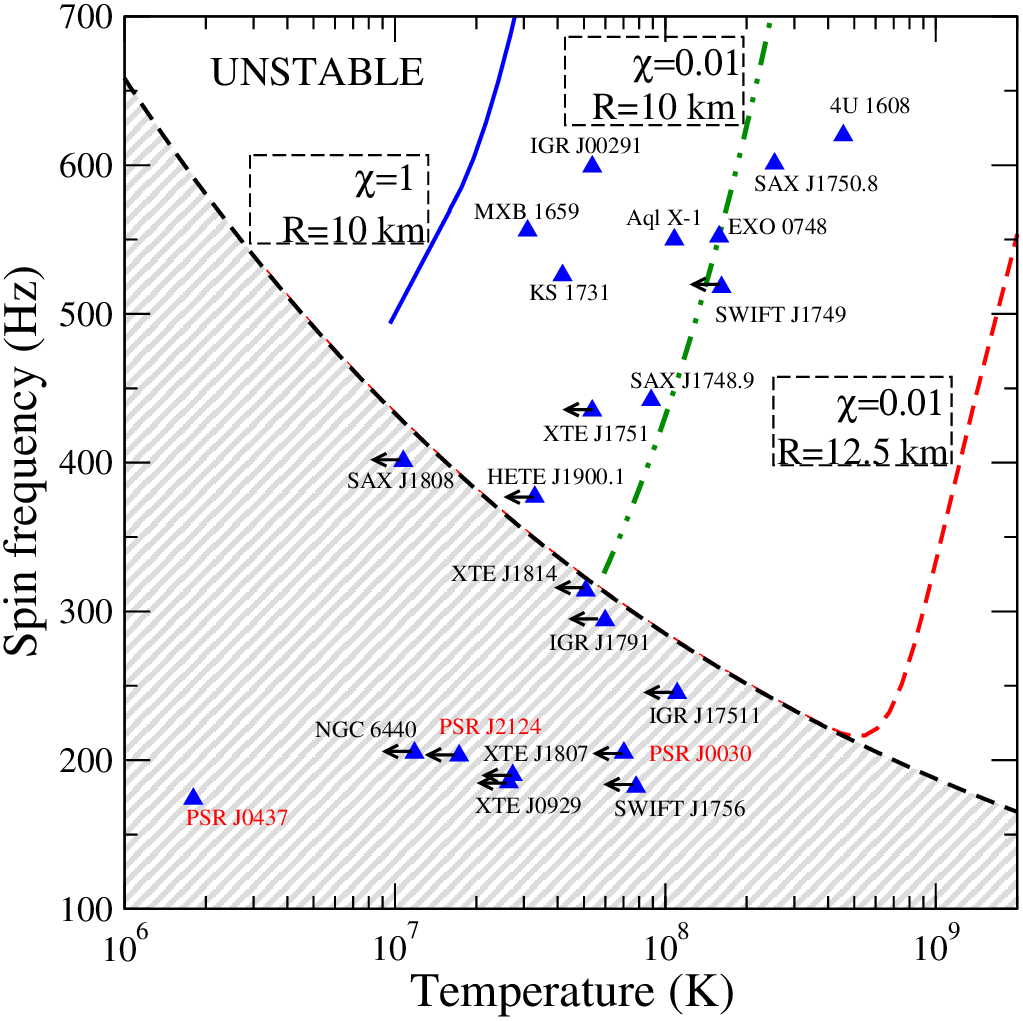}}
\caption{R-mode instability window for a $M=1.4 M_\odot$ star with hyperons in the core. We use the results of \citet{hyperon} for different NS radii and values of the coupling parameter $\chi$, which parametrises in-medium effects (see \citet{hyperon} for details). Note that the interruption at low temperatures for the $\chi=1$ curve is not physical, but due to difficulties with the numerical setup. It would appear that a rather compact star and strong coupling are required for the model to be consistent with observations.}
\label{observehyp}
\end{figure}

\begin{figure}
\includegraphics[height=7cm,clip]{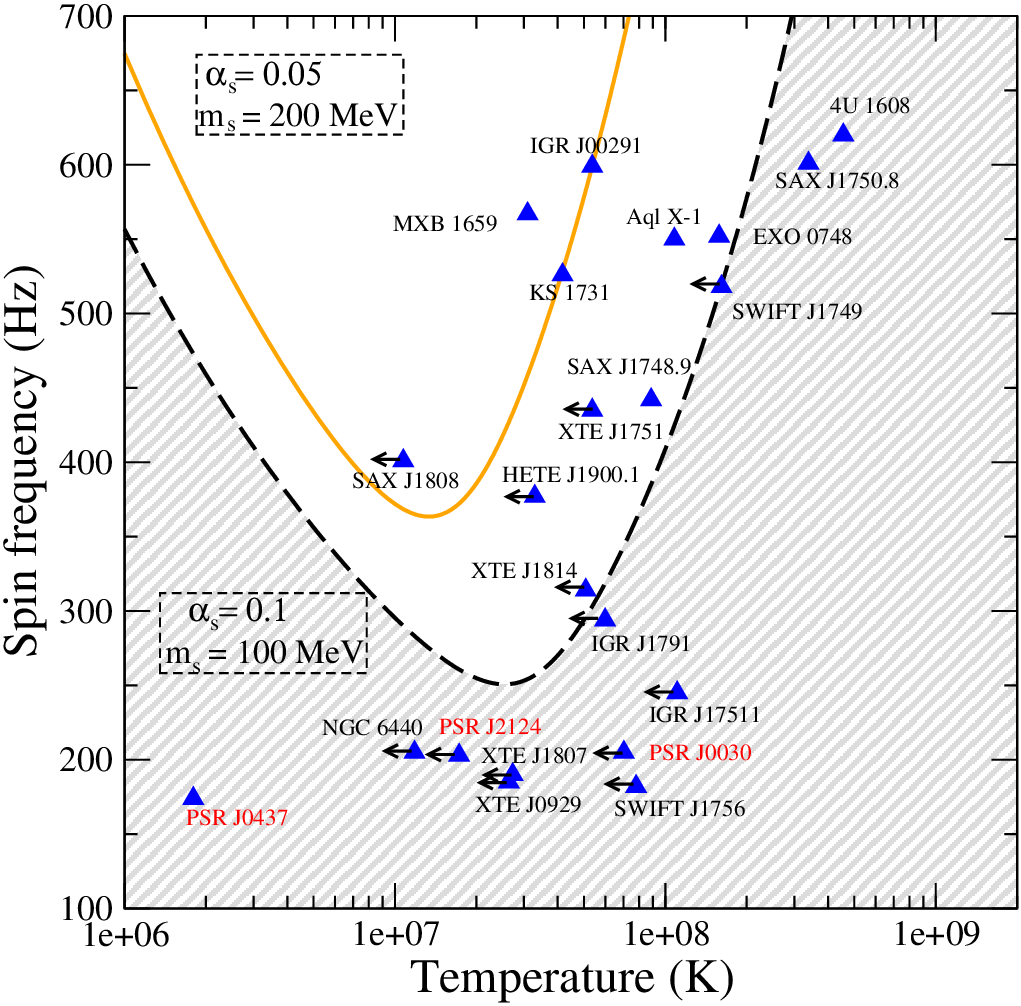}
\caption{R-mode instability window for a strange star with $M=1.4 M_\odot$ and $R=10$ km. We assume unpaired quarks in the core (see text). The observed temperature would require the strong coupling constant to be smaller than generally assumed and a rather large strange quark mass.}
\label{observestrange}
\end{figure}

Another possibility is that core bulk viscosity may be much stronger at low
temperatures.
For example, if hyperons are present in the core, then a significantly restricted unstable region is created, as illustrated in Figure~\ref{observehyp}.
The situation for strange stars is somewhat similar, with bulk viscosity playing a much stronger role at low temperatures and leading to a reduced unstable region, as shown in Figure~\ref{observestrange} (see \citealt{Alf2}, for more detailed discussion of the r-mode instability window in strange stars and hybrid stars). Although interesting, this mechanism has several problems. First, the measurement of a NS with mass $M\approx 2M_\odot$ \citep{Jason} can already exclude some hyperonic equations of state (\citealt{Lattimer}; although the presence of hyperons in the core may be consistent with the low observed temperatures of some sources reported in \citealt{Heinke1} and \citealt{Heinke2}). Second, one cannot explain the existence of millisecond radio pulsars: after accretion has ceased, one would expect most systems to cool and pass through the unstable region, during which the NS would spin down and not maintain the high spin rates that are observed ($\nu\approx 600-700$~Hz). Finally one needs hyperon/quark bulk viscosity parameters that are somewhat extreme to reconcile with the observed temperatures and spins.

\begin{figure*}
\begin{minipage}{150mm}
\centerline{\includegraphics[height=7cm,clip]{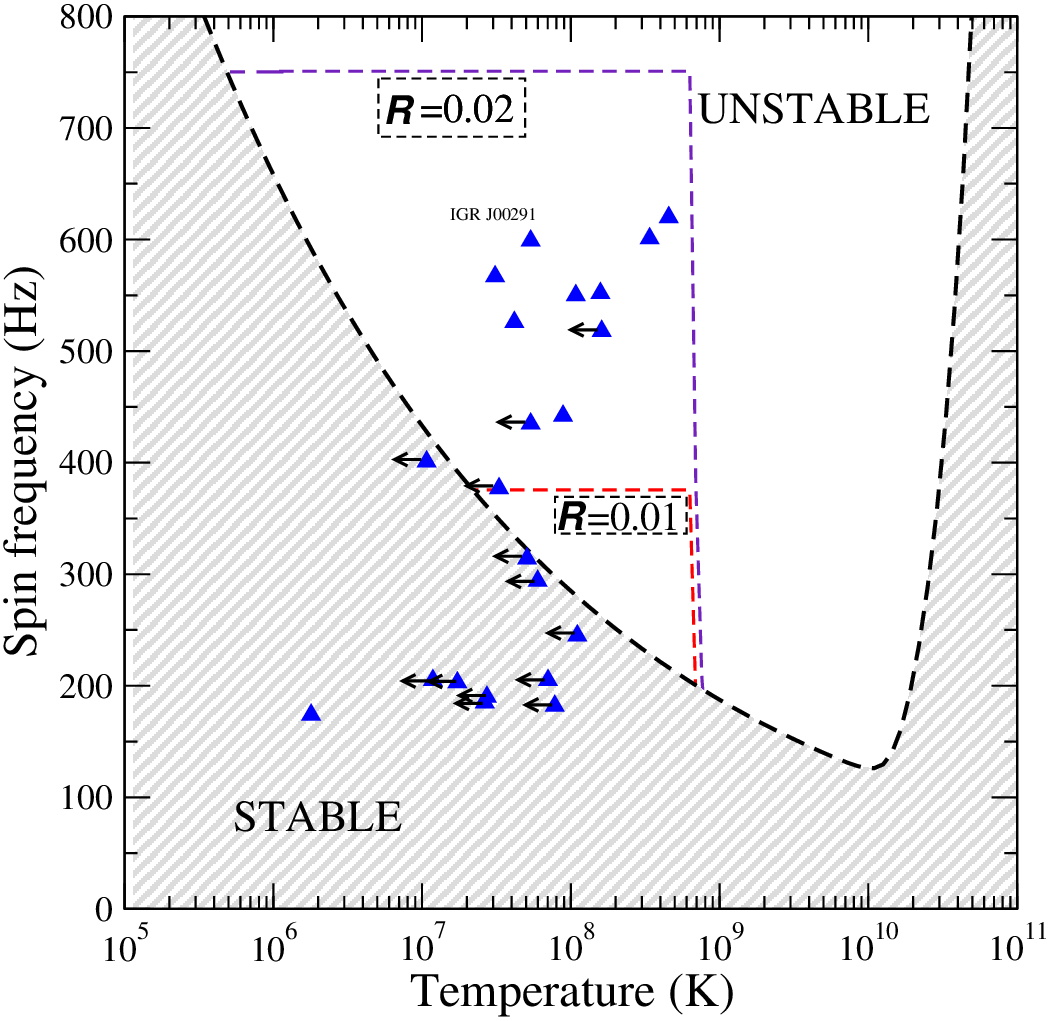}\includegraphics[height=7cm,clip]{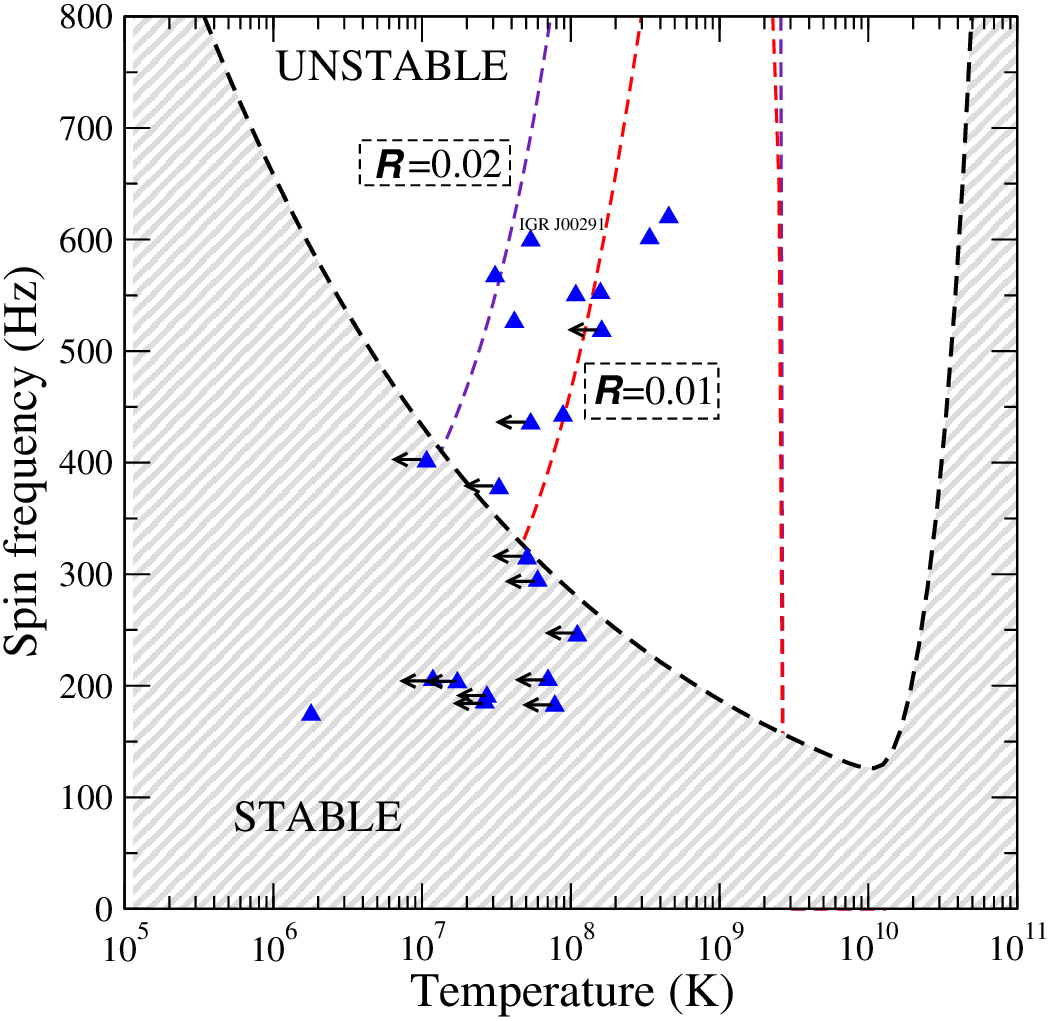}}
\caption{R-mode instability window in the presence of strong mutual friction. We use the results of \citet{rmode} for two different superfluid gap models: the so-called ``weak'' (left panel) and ``strong'' (right panel) model. In both cases, it appears that a drag parameter of the order of $\mathcal{R}\approx 10^{-2}$ is needed to explain the observations. Note that in these figures we have taken a larger temperature range, in order to better illustrate the fact that mutual friction is ineffective above the superfluid transition temperature $T_c\approx 10^9$ K.}
\label{observe2}
\end{minipage}
\end{figure*}

A very interesting mechanism is one that involves strong vortex-mediated mutual friction. If the core of the NS contains a type I superconductor, then mutual friction will not be strong enough to significantly affect the instability window (\citealt{Sedrakian}; although see \citealt{JonesI}, for a discussion of strong drag in type I superconductors). However, if the core contains a type II superconductor, then the interaction of vortices with flux tubes will lead to strong mutual friction if a large fraction of vortices can ``creep'' through the flux tubes. Examples of this are shown in Figure~\ref{observe2}.

Finally a promising scenario involves magnetic damping of the r-mode \citep{Luciano}. Given the high electrical conductivity of the NS interior, the magnetic field lines are frozen in with the fluid and can thus be distorted and wound up by the oscillatory motion of the r-mode. Even for relatively weak magnetic fields, this could lead to rapid damping and could close the instability window \citep{Luciano2, Luciano3, Cuofano}.

\section{Spin equilibrium}

We now examine the possibility that GW emission due to an unstable r-mode may be setting the spin equilibrium for LMXBs. This could be the case if the critical frequency increases with temperature at around $10^7$~K (e.g., for hyperon and quark bulk viscosity or for strong mutual friction). As a result, thermal runaway is halted, and the system reaches an equilibrium state, such that viscous heating due to the r-mode is balanced by neutrino emission and the GW torque balances the accretion torque at the observed spin period \citep{NilsST, Nayyar}.

We follow the approach of \citet{windowW} and assume that a GW torque due to an unstable r-mode is balancing a spin-up torque due to accretion on the NS surface. The heat dissipated in this case has the form \citep{BU00}
\be
L_{\mbox{heat}}=0.064\left(\frac{\nu}{300\mbox{Hz}}\right) L_{\mbox{acc}}.
\ee
Taking the heat from r-mode dissipation to be lost by neutrino emission
[i.e., $L_{\mbox{heat}}=L_\nu(T)$, where $L_\nu$ is the neutrino luminosity],
the core temperature $T$ can be inferred.
In order to determine the rate at which neutrino emission cools the system, it is important to account for superfluidity, as this will lead not only to a reduction in the emission rates for the modified Urca emission processes but also to additional neutrino emission from the formation of Cooper pairs.  We use the latest constraints on superfluid transition temperatures \citep{Page,Shternin}, obtained from the observed rapid cooling of the NS in the Cassiopea~A supernova remnant \citep{HeinkeHo,Shternin}.
We use the code described in \citet{WynnNew} to calculate the
neutrino luminosity, which is obtained by integrating the neutrino
emissivities over the stellar volume.
Briefly, this includes building a NS with the APR I equation of state
(in our case, $M=1.4\,M_\odot$ and $R=12\mbox{ km}$),
calculating neutrino emissivities due to modified Urca, nucleon
bremsstrahlung, Cooper pair formation and breaking, plasmon decay, and
pair annihilation in the core and crust,
and accounting for neutron singlet and triplet superfluids in the core
and crust, respectively, and proton superconductivity in the core.
Note that the difference between core temperatures derived here and in \citet{windowW} is $\lesssim$~five~percent and the difference between core temperatures derived using neutron triplet transition temperatures from \citet{Page} and \citet{Shternin} is $\lesssim$~fifteen~percent;
see \citet{windowW}, for derived core temperatures assuming only modified Urca neutrino emission.

\begin{figure*}
\begin{minipage}{150mm}
\centerline{\includegraphics[height=7cm,clip]{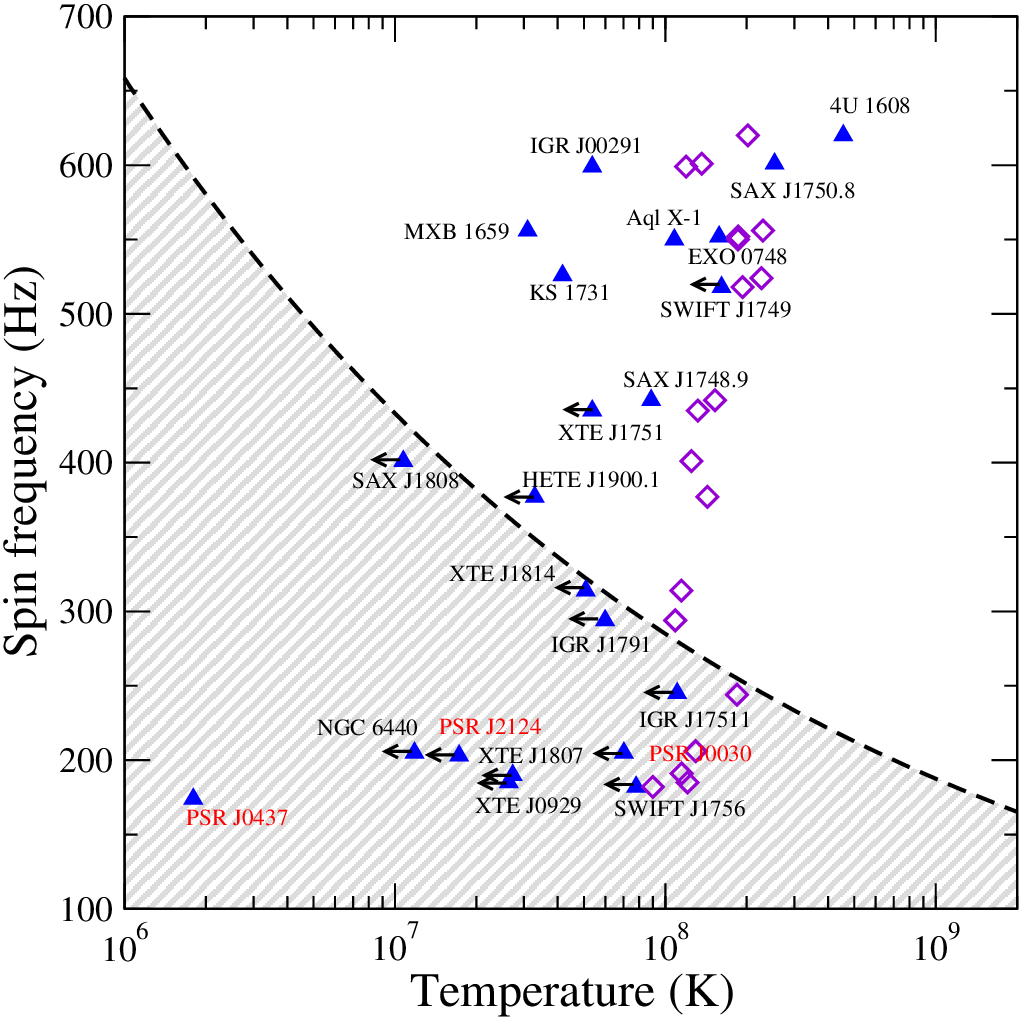}\includegraphics[height=7cm,clip]{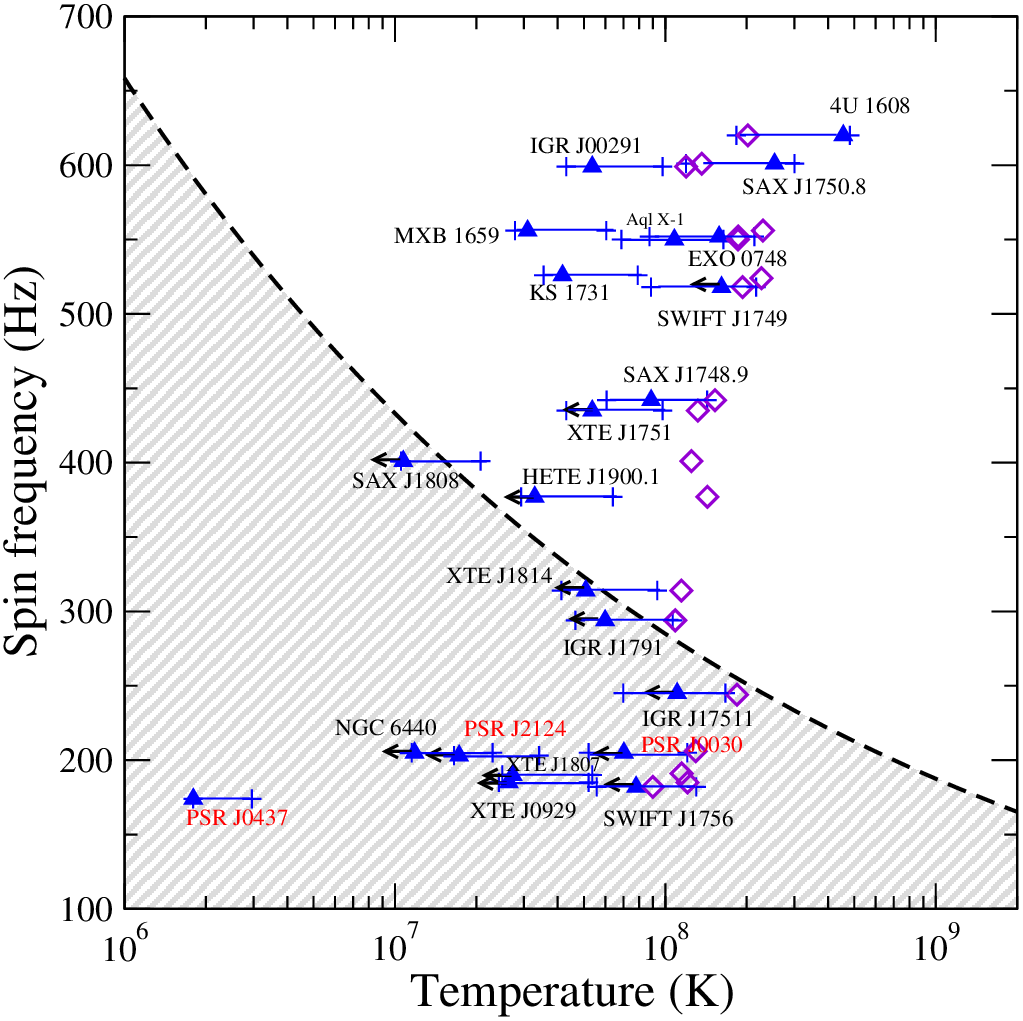}}
\caption{The left hand panel shows the core temperatures derived from X-ray luminosities of LMXBs measured during bursts and assuming spin equilibrium (diamonds),
and core temperatures derived here from UV and quiescent X-ray luminosities (triangles).
The panel on the right hand side is the same but also shows the range in temperatures obtained by assuming that the minimum core temperature is given by assuming an envelope completely composed of light elements, while the maximum is for a pure iron envelope, as described in the text.
It is obvious that many systems are too cold to allow for a spin equilibrium r-mode.  However, the more rapidly rotating systems, which are hotter, may be consistent with spin equilibrium.}
\label{error}
\end{minipage}
\end{figure*}

In Figure~\ref{error}, we show the temperatures obtained in the spin equilibrium scenario for a ``shallow'' neutron superfluid transition (see \citet{WynnNew} for details) with $T_{\mathrm{cn,max}}\approx 5\times 10^8$~K, as in \citet{Page}. The long-term accretion luminosities are taken from \citet{Watts08} and from \citet{Falanga} for IGR J17511-3057. We can see that many systems appear to be colder than what would be expected in the presence of an unstable r-mode, although for some of the faster systems (which are also the most likely targets for GW searches, given the strong scaling with frequency of the GW torque), GW-driven spin equilibrium may still be possible and cannot be completely ruled out. 

%\begin{figure}
%\centerline{\includegraphics[height=7cm,clip]{errorsblue.eps}}
%\caption{Spin equilibrium temperature}
%\label{error}
%\end{figure}

Finally we can calculate the maximum amplitude that would be compatible with the inferred core temperatures by assuming that the viscous heating is due to an unstable r-mode with arbitrary amplitude $\alpha$ \citep{review}
\be
L_{\mbox{heat}}=1.31\frac{\alpha^2\nu^2 M R^2}{\tau_{sv}},
\ee
where $M$ is the mass of the star, $R$ its radius, we have assumed that the equation of state is given by an $n=1$ polytrope and $\tau_{sv}$ is the shear viscosity damping timescale. If we assume that shear viscosity is mainly due to electron-electron scattering and that modified Urca reactions are the main contribution to the cooling, the maximum amplitude we obtain takes the form
\be
\alpha_{m}\approx 6.7\times 10^{-5}\frac{T_8^5}{\nu}
\ee
where $T_8$ is the temperature in units of $10^8$ K and $\nu$ is the spin frequency of the system.
With the notable exception of the two fastest and hottest systems, 4U 1608-52 and SAX J1750.8-2900, these are more stringent upper limits on the mode amplitude than the spin equilibrium condition and from the results in table \ref{tempspin} we see that for most systems in the unstable region one has $\alpha_m\approx 10^{-9}-10^{-8}$, which could be compatible with a small saturation amplitude for the r-mode and would have no impact on the spin evolution of the system \citep{Bondarescu}.

\section{Conclusions}

In this paper, we estimated the core temperature of NSs using data from X-ray observations of LMXBs in quiescence and UV observations of millisecond pulsars, in order to place constraints on the physics of the r-mode instability window.
We also presented a new analysis of five systems.

These estimates show that, if one uses a  ``minimal'' NS model, in which shear viscosity is due to dissipation in a boundary layer between the crust and core and bulk viscosity is due to modified Urca processes, the r-mode would be unstable in many of these systems. In particular, many systems are well above the critical frequency-temperature for the instability to grow, which is highly unlikely since systems should never depart significantly from the stable region for small saturation mode amplitudes (such as those predicted by \citealt{Bondarescu}), while systems should spin down too quickly to be detected for much larger values of the saturation amplitude \citep{Heyl02}.

It is clear that additional physics and additional damping mechanisms have to be built into the model for it to be consistent with observations. Enhanced bulk viscosity, due to hyperons or deconfined quarks in the core, could provide a source of damping that is consistent with observations, but their presence would also predict that, as systems cool after accretion ceases, they should once again enter an unstable region and spin down.  This is at odds with the existence of rapidly rotating millisecond radio pulsars.
Strong mutual friction may also be consistent with observations if superfluid vortices can cut through superconducting flux tubes (see Glampedakis, Andersson \& Haskell, in preparation, for a detailed discussion of this scenario).

A promising scenario is that in which damping is due to the crust responding rigidly to the mode displacement in the r-mode frequency range for rapidly rotating system. The phenomenological model of \citet{windowW} shows that this is viable, but more quantitative models are needed. In particular, efforts should be made to better understand the effect on viscous damping timescales of pasta phases at the crust-core interface \citep{HB08} .
Another scenario that would be consistent with observations is that in which the mode winds up the magnetic field of the star, and energy is extracted from the oscillatory motion more rapidly than GW emission can drive it \citep{Luciano}. Once again this mechanism depends strongly on the internal magnetic field structure and further work is needed in order to assess its relevance, as well as accounting for the presence of superconducting components.

Finally an interesting possibility is that the saturation amplitude of the r-mode is small enough that the GW torque cannot counteract the accretion torque and a system would spin up into the unstable region. In order for this scenario to be consistent with observations (i.e. in order for the heating from the mode to be consistent with the observed temperature), the saturation amplitude should be roughly $\alpha\la 10^{-9}-10^{-8}$. Such a small amplitude may be consistent with theoretical calculations \citep{Bondarescu} and would indeed lead to a spin-down torque that is smaller than the electromagnetic spin-down torque for a $B\approx 10^8$ G magnetic field, thus not impacting on the evolution of the systems.

We examined the possibility that continuous GW emission from an unstable r-mode may be setting the spin equilibrium period of the LMXBs. This scenario was considered by \citet{BU00} who found that, if one assumes modified Urca cooling, most systems would be too hot to be consistent with observations. We re-examined this scenario by using the most recent constraints on superfluid transition temperatures obtained from observations of the cooling of the NS in Cassiopeia~A. We find that this leads to lower core temperatures (due to stronger neutrino emission) which may be consistent with the more rapidly rotating systems. This is interesting since the GW spin-down torque scales strongly with frequency and is expected to play a stronger role in rapidly rotating systems.
Further observational constraints, as may be available from future X-ray observatories such as LOFT and Astrosat, as well as theoretical work on NS composition and viscosity, are crucial to aid in the search for GWs from these systems \citep{Watts08}.

\section*{Acknowledgments}
BH acknowledges support from the European Union via a Marie Curie IEF fellowship.ND is supported by the Netherlands organisation for scientific research (NWO) and by NASA through Hubble Postdoctoral Fellowship grant number HST-HF-51287.01-A from the Space Telescope Science Institute, which is operated by the Association of Universities for Research in Astronomy, Incorporated, under NASA contract NAS5-26555.
WCGH appreciates the use of the computer facilities at KIPAC.
WCGH acknowledges support from STFC in the UK.
This work made use of the public data archive of Swift

\end{document}